 \documentclass[final,5p,times,twocolumn,authoryear]{elsarticle}

\usepackage{amssymb}
\usepackage{lipsum}
\usepackage{amsmath}
\usepackage{cancel}

\journal{arXiv}

\begin{document}

\begin{frontmatter}



\title{Factorization of Coupling Constant in Background Field}


\author[first]{Cong Li}
\affiliation[first]{organization={Zhejiang Ocean University},
            addressline={congli@zjou.edu.cn}, 
            city={Zhoushan},
            postcode={316022}, 
            state={Zhejiang},
            country={China}}

\begin{abstract}
Due to color confinement, hadrons are inseparable when studying strong interactions. The paper mainly studies the scattering in the background field, especially the scattering of quarks in the hadronic background field. We discovered that the coupling constant within the background field is made up of the distribution function of background field. The finding provides a foundational comprehension of coupling constants, which significantly contributes to our understanding of running coupling constant and the perturbative nature of theory.
\end{abstract}




\end{frontmatter}




\section{Introduction}
\label{introduction}

Due to color confinement, hadrons are inseparable in the study of strong interactions, necessitating the exploration of scattering within the hadronic background field. While study on scattering processes typically emphasizes multiple scattering effects \cite{schwartz}, the paper studies additional impacts of the background field on the scattering process.

Our current understanding of the coupling constant encompasses two key aspects: it signifies the strength of interaction and runs with energy scale. In the study, we incorporate the background field impacts into the coupling constant, akin to factoring the coupling constant. We find that the coupling constant is equivalent to the distribution function of the background field. It shows that the ability of electrons to exchange photons with momentum $k$ depends on the number of photons with momentum $k$ in the background field. This discovery uncovers the meaning of the coupling constant, rather than making corrections to it.

The paper consists of four sections. The second section focuses on the scattering of electrons in the photon background field and the factorization of electromagnetic coupling constants in the background field. The third section studies the coupling constant of strong interactions and provides a specific meaning of it. The fourth section primarily introduces some distinctive coupling constants. Finally, a straightforward summary concludes the article.
\begin{figure}
	\centering 
	\includegraphics[width=0.13\textwidth, angle=0]{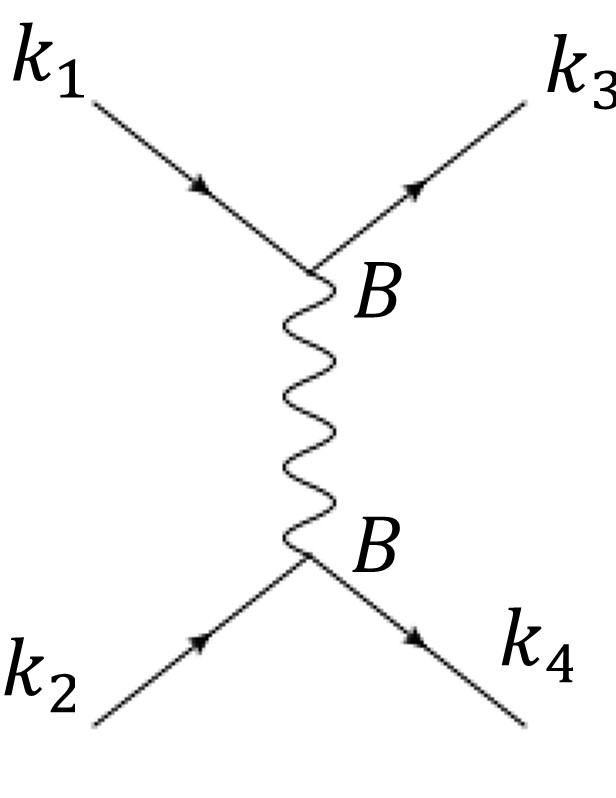}	
	\caption{Electron-Electron scattering in $B$} 
	\label{1}%
\end{figure}

\section{Factorization of electromagnetic coupling constant in background field}

First, electron-electron scattering is taken as an example to study photon propagators in the photon background field. The specific reaction can be represented by the Figure 1, where $B$ represents the background field. Its amplitude before Wick contraction is
\begin{equation}
    \left\langle k_3,k_4,B\middle|\bar{\psi}\left(x\right)A_\mu\left(x\right)\psi\left(x\right)\bar{\psi}\left(y\right)A_\nu\left(y\right)\psi\left(y\right)\middle| k_1,k_2,B\right\rangle,
\end{equation}
where we are concerned with the contraction of the photon field. Contract it with the photon background field $B$ \cite{peskin}, We get
\begin{equation}
    \left\langle B\middle| A_\mu\left(x\right)A_\nu\left(y\right)\middle| B\right\rangle.
\end{equation}
Structurally, it is a propagator, but distinguished from multiple scattering and photon correlation function. We directly use Wick contraction to deal with equation (2), after inserting two $1=\int{\frac{d^3k}{\left(2\pi\right)^32k_0}\left|k\right\rangle \left\langle k\right|}$ into equation (2). Then
\begin{equation}
    \begin{aligned}
& \left\langle B\left|A_\mu(x) A_v(y)\right| B\right\rangle \\
& =\int \frac{d^3 k}{(2 \pi)^3 2 k_0} \int \frac{d^3 k^{\prime}}{(2 \pi)^3 2 k_0^{\prime}}\langle B \mid k\rangle\left\langle k\left|A_\mu(x) A_v(y)\right| k^{\prime}\right\rangle\left\langle k^{\prime} \mid B\right\rangle \\
& =\int \frac{d^3 k}{(2 \pi)^3 2 k_0}\langle B \mid k\rangle\left\langle k\left|A_\mu(x) A_v(y)\right| k\right\rangle\langle k \mid B\rangle  \\
& =\int \frac{d^3 k}{(2 \pi)^3 2 k_0}|\langle B \mid k\rangle|^2 \mathrm{e}^{-i k \cdot(x-\mathrm{y})} g_{\mu v} \\
& =\int \frac{d^3 k}{(2 \pi)^3 2 k_0}|\langle B \mid k\rangle|^2 \mathrm{e}^{-i k \cdot(x-\mathrm{y})} g_{\mu v}  \\
& =\int \frac{d^4 k}{(2 \pi)^4} \mathrm{e}^{-i k \cdot(x-\mathrm{y})} \frac{-\mathrm{i} g_{\mu v}}{k^2+\mathrm{i} \varepsilon}|\langle B \mid k\rangle|^2.
\end{aligned}
\end{equation}
The last equation uses the inverse process of contour integral. It should be noted that, the photon is no longer on shell at this time, and the on shell condition is absorbed into the integral of energy. In the second equation, the integral of $d^3k^\prime$ is removed due to $\left\langle k\middle| k^\prime\right\rangle=\delta\left(k-k^\prime\right)$, only the term $\left\langle k\middle| A_\mu\left(x\right)A_\nu\left(y\right)\middle|k\right\rangle$ can survive. The third equation uses Wick contraction. As you can see, the photon propagator in the background field is different from the photon propagator in the vacuum. And it is the propagator in the vacuum convolved with a distribution function $\left|\left\langle B\middle| k\right\rangle\right|^2=f\left(k\right)$, which represents the probability density distribution of on shell photons in the background field. The propagator is also true for other bosons. For a more complex photon background field, its partons may be polarized, and we can also write its propagator
\begin{equation}
    \int{\frac{d^4k}{(2\pi)^4}e^{-ik\cdot\left(x-y\right)}\frac{-i}{k^2+i\varepsilon}}\left[g_{\mu\nu}f_1\left(k\right)+\left(g_{\mu\nu}-2k_\mu k_\nu/k^2\right)f_2\left(k\right)\right],
\end{equation}
where $f_1$ and $f_2$ represent the non-polarized and linlearly polarized photon distribution functions respectively. When it is the fermion background field, the propagator is
\begin{equation}
    \int \frac{d^4 p}{(2 \pi)^4} \mathrm{e}^{-i p \cdot(x-\mathrm{y})} \frac{i(\cancel{p} +m) f(p)}{p^2-\mathrm{m}^2+i \epsilon},
\end{equation}
where the $f\left(p\right)$ represents the on-shell fermion distribution function.

For equation (3), we can absorb the distribution function into the coupling constant and obtain the effective coupling constant $\alpha_{eff}\left(k\right)$, which consists of the coupling constant $\alpha_e$ and the distribution function of the background field $f\left(k\right)$. For electron-electron scattering in the photon background field, we have
\begin{equation}
    \alpha_{eff}\left(k\right)=\alpha_ef\left(k\right),
\end{equation}
where $\alpha_e$ is the electromagnetic coupling constant $\frac{1}{137}$, and $\alpha_{eff}(k)$ represents the experimentally measured electromagnetic interaction intensity in this background field. As you can see from equation (6), the effective coupling constant is affected by the photon distribution function in the background field. When we take the derivative of both sides of the equation (6) with $Q^{2}=-k^{2}$ and multiply by $Q^{2}$, we have
\begin{equation}
    Q^2\frac{d\alpha_{eff}\left(k\right)}{dQ^2}=\alpha_eQ^2\frac{df\left(k\right)}{dQ^2}\ +Q^2\ \frac{d\alpha_e}{dQ^2}f\left(k\right).
\end{equation}
The left side of the equation indicates the running of the effective coupling constant, and the first term on the right side indicates the scaling evolution of the photon distribution function in the background field, while the second term represents the running of the coupling constant in vacuum.

\section{Factorization of coupling constant for strong interactions in nucleons}

Based on the similarity of gauge theory, we continue to study the strong interaction coupling constant $\alpha_s$ this section. We commonly use the proton-proton collision to measure the coupling constant of strong interactions. When we extract the coupling constant $\alpha_s$ from the experimental data, the coupling constant is the effective coupling constant $\alpha_s\left(k\right)$ in fact. Because quark-quark scattering occurs in the background field, which is protons, a complex system of quarks and gluons. Analogy equation (6), we have
\begin{equation}
   \alpha_s^{eff}\left(k\right)=\alpha_sg\left(k\right),
\end{equation}
where $g\left(k\right)$ is the distribution function of gluons within the proton, and $\alpha_s=\frac{g^2}{4\pi}$. In fact, the effective coupling constant is also the coupling constant we are familiar with for strong interactions in theory. After derivation and multiplying by $Q^2$, it is
\begin{equation}
   Q^2\frac{d\alpha_s^{eff}\left(k\right)}{dQ^2}=\alpha_sQ^2\frac{dg\left(k\right)}{dQ^2}\ +Q^2\ \frac{d\alpha_s}{dQ^2}g\left(k\right).
\end{equation}
In addition to the running of the coupling constant, the most noticeable thing is $\alpha_s Q^2\frac{dg\left(k\right)}{dQ^2}$. In protons, the running of the strong interaction coupling constant and the DGLAP evolution equation of the gluon distribution function have been fully studied and are in good agreement with experiments \cite{dglap} \cite{coupling1} \cite{coupling2}. Put the two evolution equations into equation (9), we find that $Q^2\ \frac{d\alpha_s}{dQ^2}g\left(k\right)=0$. 

The proof of $Q^2\ \frac{d\alpha_s}{dQ^2}g\left(k\right)=0$ is as follows. Firstly, we choose small $x$ protons, where $x$ is the momentum fraction of parton in the proton. The components of the proton are very simple, all are gluons, and quarks can almost be ignored. By expand the running coupling constant and distribution function to $\alpha_s$ order, we can obtain the differentials of the gluon distribution function and coupling constant respectively, as shown in Figure 2 and Figure 3. 
 \begin{figure}
	\centering 
	\includegraphics[width=0.4\textwidth, angle=0]{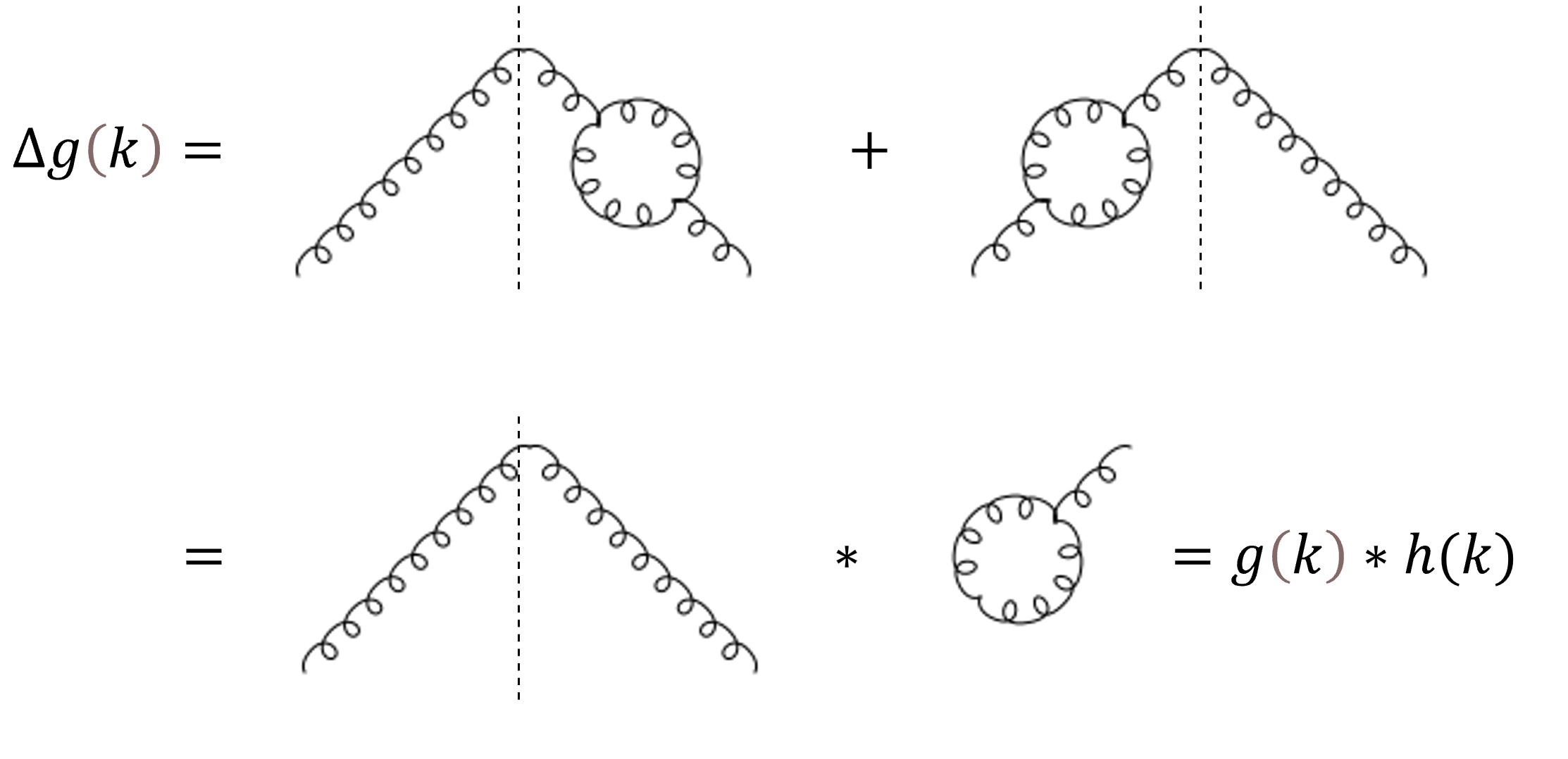}	
	\caption{Differential of the gluon distribution function} 
	\label{2}%
\end{figure}
  \begin{figure}
	\centering 
	\includegraphics[width=0.42\textwidth, angle=0]{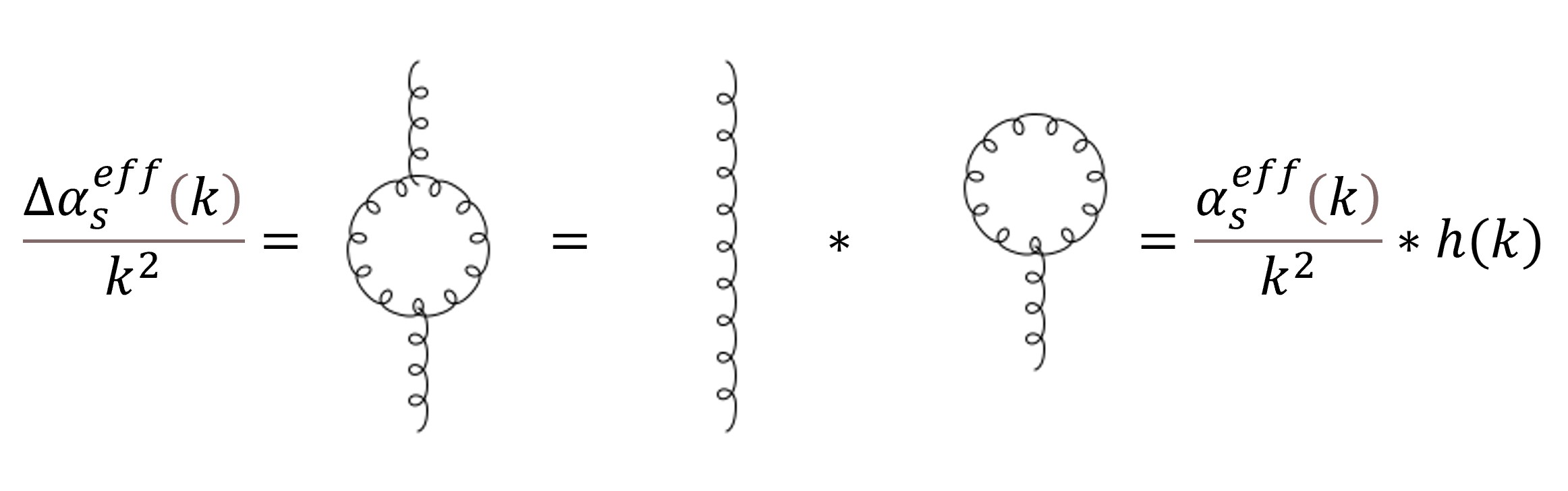}	
	\caption{Differential of coupling constant} 
	\label{3}%
\end{figure}
Where $*$ represents convolution, and the contribution of the relevant real diagram is not listed because its contribution is mathematically equivalent to the virtual diagram \cite{dglap}. $h\left(k\right)$ is part of the hard process, which is a gluon loop with a leg. In the same background field, $h\left(k\right)$ can be perturbed and calculated, and $h\left(k\right)$ is the same in both cases. From the figure, we get the following two equations,
\begin{equation}
   \Delta g\left(k\right)=g\left(k\right)\ast h\left(k\right),
\end{equation}
\begin{equation}
   \Delta \alpha_s^{eff}\left(k\right)=\alpha_s^{eff}\left(k\right)\ast h\left(k\right).
\end{equation}
For equation (10), multiplying both sides by $\alpha_s$, we have
\begin{equation}
   \alpha_s\Delta g\left(k\right)=\mathrm{\Delta}\alpha_s^{eff}\left(k\right).
\end{equation}
Comparing equation (9) with equation (12), it is easy to find that $Q^2\ \frac{d\alpha_s}{dQ^2}f\left(k\right)=0$, which shows that $\alpha_s$ here no longer runs. This is not difficult to understand. It can be seen from equation (6) that $\alpha_s$ represents the coupling constant of the strong interaction in vacuum. In the renormalization group equation, its running mainly comes from $h\left(k\right)$ of gluons in vacuum. When the background is protons, the running of this coupling constant is mainly due to the $h\left(k\right)$ of the gluons in the protons. In the same way, the coupling constant $\alpha_s$ is also rapidity-independent. 

In summary, $\alpha_s$ at this time is a constant used to balance the relationship between the distribution function and the coupling constant.
The running of the coupling constant is extracted completely. The reason for the running is the scaling evolution of the parton distribution function. Using $\alpha_s\int{g\left(k\right)\ }\frac{d^3k}{\left(2\pi\right)^32k^0}=\alpha_sn=N$, where $n$ is the number of gluons and $N$ is the effective number of gluons, the constant $\alpha_s$ can be absorbed into the distribution function. We have
\begin{equation}
   \alpha_s^{eff}\left(k\right)=g\left(k\right).
\end{equation}
As you can see, the strength of the strong interaction depends on the specific number of gluons with momentum $k$ in the background field. The more gluons, the more energy and momentum are transferred. It is worth mentioning that the coupling constants measured for proton-proton collisions and proton-neutron collisions should be different. Because the gluon distributions are different. Of course, the difference will not be very large, because the distribution functions of gluons in protons and neutrons themselves are not very different.

The factorization of coupling constants in protons was discussed above. But for the coupling constant in vacuum, it is not yet known whether it is equivalent to the distribution function of virtual gluons in vacuum. The above study shows that the gluons in the background field are responsible for transferring energy and momentum, and the distribution function reflects the efficiency of transferring energy and momentum. We can boldly equate the coupling constant in vacuum to the distribution function of virtual gluons (or photons) in vacuum. It would be studied in the future.

\section{Other situations}

As mentioned above, in hadrons, the coupling constant of the strong interaction is equivalent to the distribution function of gluons in hadrons. In some complex processes, the relationship between the coupling constant and the distribution function may be more complex. In Electron-Ion Collision, since electrons have no color, we usually use quark Dipole to detect the distribution of gluons within hadrons. This process can be represented by Figure 4. The electron emits a virtual photon, which fluctuates into quark pairs, and the quark pairs scatter with the gluon field. There are two special propagators here, which are on-shell quark propagators, represented by fermion lines with cut. The on-shell propagator in the background field can be written as
\begin{equation}
   (\cancel{p}+m) q(p) \mathrm{e}^{-i p \cdot(x-y)},
\end{equation}
where $q\left(p\right)$ is the distribution function of quarks within the hadron. At this time, the coupling constant of electromagnetic interaction is not only affected by the photon field within the hadron, but also affected by the quark distribution function within the hadron. Regardless of the quark's transverse momentum, the product of all coupling constants in the quark's center-of-mass system is
\begin{equation}
\alpha_e^{total}\left(k\right)=\alpha_e^2f^2\left(k\right)q\left(k/2\right)\bar{q}\left(k/2\right),
\end{equation}
where $\bar{q}$ represents the distribution function of antiquark. Assuming that the distribution of quarks and antiquarks is exactly the same, we have
\begin{equation}
   \alpha_e^{eff}\left(k\right)=\alpha_ef\left(k\right)q\left(k/2\right).
\end{equation}
\begin{figure}
	\centering 
	\includegraphics[width=0.4\textwidth, angle=0]{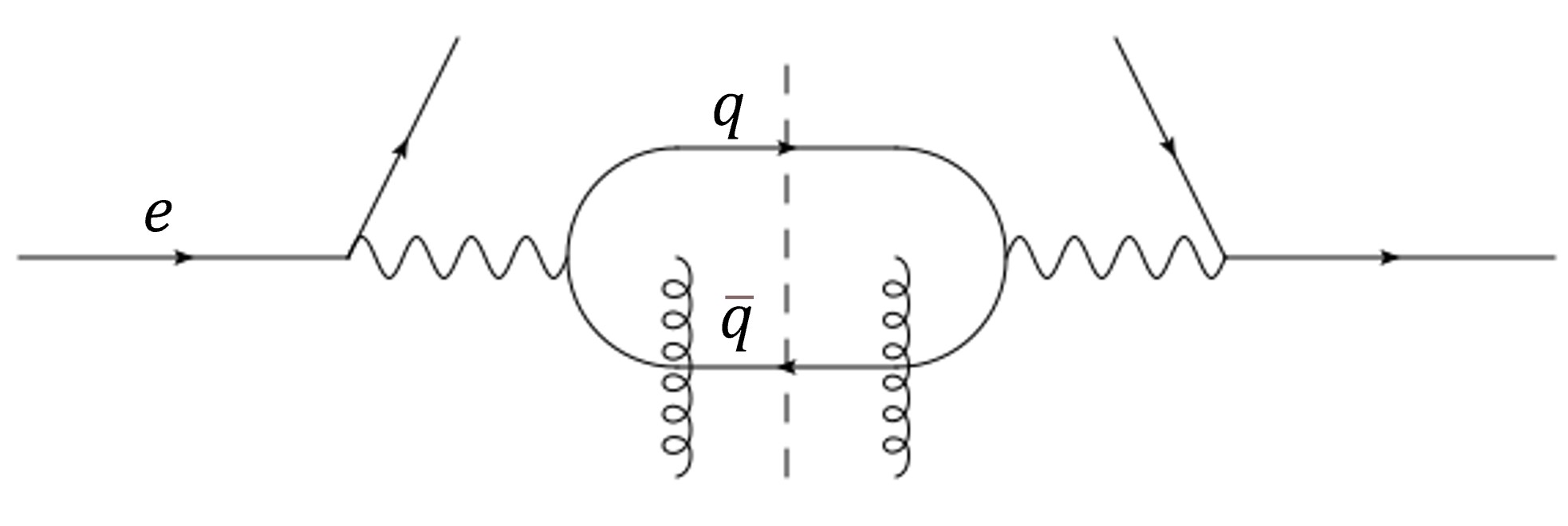}	
	\caption{Scattering of electron-ion} 
	\label{4}%
\end{figure}
As you can see, the coupling constant is affected not only by the photon distribution function, but also by the quark distribution function. It can be seen that for a similar process, although the density of gluons is large, the probability density of quarks is small, the coupling constant of the process may still be small, and the perturbation theory is still valid. At this time, the running of the coupling constant is also affected by the scaling evolution of the two distribution functions
\begin{equation}
  Q^2\frac{d\alpha_e^{eff}\left(k\right)}{dQ^2}=\alpha_eQ^2\frac{df\left(k\right)}{dQ^2}q\left(k/2\right)\ +\alpha_eQ^2\ \frac{dq\left(k/2\right)}{dQ^2}f\left(k\right).
\end{equation}

\section{Summary and outlook}
The paper studies the impact of background fields on the scattering process, highlighting how the presence of the fields alters the propagators, distinguishing it from scattering in a vacuum. Our approach involves integrating the distribution function into the coupling constant to derive the effective coupling constant, effectively leading to what we term as the "factorization of the coupling constant." Our study reveals that the running of the effective coupling constant comes from the scaling evolution of the distribution function, providing a novel interpretation of the coupling constant as the probability density function of bosons within the background field. It also elucidates why the effect of background fields could be previously overlooked, due to running of coupling constant. The paper also briefly studies the coupling constants of more intricate processes, emphasizing their reliance on multiple distribution functions and the determination of their strength. 

In the future, we focus on resolving the relationship between virtual particle distribution functions in vacuum with coupling constants in vacuum.

\section*{Acknowledgements}
Thanks to Du-xin Zheng for helpful discussions.

\bibliographystyle{elsarticle-harv} 
\bibliography{example}






\end{document}